\documentclass[]{jfm}
\usepackage{graphicx}
\usepackage{amsmath}
\usepackage{multirow}
\usepackage{color}

\newcommand{\bm}[1]{\boldsymbol{#1}}
\newcommand{\dd}[2]{\frac{\partial #1}{\partial #2}}
\newcommand{\figref}[1]{figure \ref{#1}}

\newcommand{\Kair}{\color{black}{}}

\graphicspath{{./}}

\shorttitle{Nonlinear mode decomposition for fluid dynamics}
\shortauthor{T. Murata, K. Fukami and K. Fukagata}

\title{Nonlinear mode decomposition with {\Kair{convolutional neural networks}} for fluid dynamics} 
\author{Takaaki Murata\aff{1},
  Kai Fukami\aff{1,2}
 \and Koji Fukagata\aff{1}\corresp{\email{\color{black}fukagata@mech.keio.ac.jp}}}

\affiliation{\aff{1}Department of Mechanical Engineering, Keio University, Yokohama, 223-8522, Japan
\aff{2}Department of Mechanical and Aerospace Engineering, University of California, Los Angeles, CA 90095, USA}

\begin{document}

\maketitle

\begin{abstract}
We present a new nonlinear mode decomposition method 
to visualize the decomposed flow fields, named the mode decomposing convolutional neural network {\color{black} autoencoder (MD-CNN-AE)}.
The proposed method is applied to a flow around a circular cylinder at $Re_D=100$ as a test case.
The flow attributes are mapped into two modes in the latent space and then these two modes are visualized in the physical space.  
{Because the \color{black} MD-CNN-AEs} with nonlinear activation functions show lower reconstruction errors than the proper orthogonal decomposition (POD), the nonlinearity contained in the activation function is considered the key to improve the capability of the model.
It is found by applying POD to each field decomposed using the {\color{black} MD-CNN-AE} with hyperbolic tangent activation  that a single nonlinear {\color{black} MD-CNN-AE} mode contains multiple orthogonal bases, in contrast to the linear methods, i.e., POD and {\color{black} the MD-CNN-AE} with linear activation.
{\Kair We further assess the proposed MD-CNN-AE by applying it to a transient process of a circular cylinder wake in order to examine its capability for flows containing high-order spatial modes.}
The present results suggest a great potential for the nonlinear {\color{black} MD-CNN-AE} to be used for feature extraction of flow fields in lower dimension than POD, while retaining interpretable relationships with the conventional POD modes.

\end{abstract}

\begin{keywords}
low-dimensional models, vortex shedding, computational methods
\end{keywords}

\vspace{-15mm}
\section{Introduction}

Mode decomposition methods have been desired to understand the physics of complicated fluid flow phenomena containing high nonlinearity and chaotic nature.
Proper orthogonal decomposition (POD)  \citep{lumely1967} and dynamic mode decomposition (DMD) \citep{Schmid2010} are well-known methods for reduced order modelling, which efficiently extract low dimensional modes. 
With both methods, the key structures embedded in the time series of flow fields can be found and visualized, although there is a difference in the sense that POD determines the optimal set of modes to represent data based on energy norm, while DMD captures dynamic modes with associated growth rates and frequencies \citep{taira2017}.
These methods have helped us to understand the important structures underlying flow phenomena and to compare flow fields under different conditions \citep{Murray2009}.
In addition, it is possible to construct control laws based on reduced order models at low computational costs \citep{bergmann2005,SAMIMY2007,Rowley2017}, since the time-evolving flow field can be represented by a linear combination of the expansion coefficients and the orthogonal bases.  
However, it is not easy to deal with highly nonlinear problems, such as high Reynolds number flows, using the conventional reduced order models  because of their linear nature.
With POD, for example, 7260 modes are necessary to reproduce 95\% of total energy for a turbulent channel flow at $\Rey_\tau=180$ \citep{alfonsi2006}, while we need only two POD modes to reproduce 99\% of total energy for a flow around a circular cylinder at $\Rey_D=100$.
This limitation narrows the applicability of the conventional reduced order models to various flow fields.

In recent years, machine learning has been widely applied in the field of fluid dynamics, and is highly expected for its strong ability to account for nonlinearity \citep{BruntonNoack2015,Kutz2017,taira2019,Brunton2019}.
\citet{LKT2016} used a customized multi-layer perceptron accounting for the Galilean invariance for Reynolds-averaged Navier--Stokes turbulence modelling. 
For large-eddy simulation, \citet{MS2017} used a multi-layer perceptron to estimate the eddy viscosity with the blind deconvolution method.  
The recent efforts for turbulence modelling are summarized well in \citet{DIX2019}.  
Machine learning has been also utilized for reduced-order modeling.  
\citet{SM2018} proposed the extreme learning machine based reduced-order modelling for turbulent systems and showed its advantage against POD.
The multi-layer perceptron and long short term memory are utilized for developing the temporal evolved turbulence with the nine-equation shear flow model by \citet{SGASV2019}.  
In this way, the fusion of machine learning and fluid dynamics is ongoing now.

Especially, the convolutional neural network (CNN) \citep{Lecun1998}, widely used for image processing, has been utilized as an appropriate method to deal with flow field data with the advantage that we can handle the fluid {\it big data} with reasonable computational costs thanks to the concept of filters called weight sharing.  
\citet{Fukami2019a} performed a super-resolution analysis for two-dimensional turbulence using a customized CNN to adapt multi-scale phenomena.
The deep CNNs were also considered to predict the small scale in the ocean turbulence called `atoms' by \citet{SP2019}.  Of particular interest of CNN is the application for dimension reduction via autoencoder \citep{hinton2006}.
{\Kair The autoencoder composed of linear perceptrons is known to work similarly to POD \citep{BH1989}.
For applications to fluid mechanics, too, \citet{Milano2002} have successfully demonstrated through various problems, such as randomly forced burgers equation and turbulent channel flows, that the capability of autoencoder is significantly improved by adopting nonlinear multi-layer perceptrons.
In addition, autoencoders have recently exhibited their remarkable ability in combination with CNN not only in the field of image processing but also in fluid mechanics.} 
\citet{omata2019} proposed a method utilizing a CNN autoencoder with POD to reduce the dimension of two-dimensional airfoil flow data.  
These concepts have also been applied to develop an inflow turbulence generator by \citet{Fukami2019b}.  
Despite these favorable properties, the conventional CNN autoencoders are interpretable only in terms of the input, the latent vector (i.e., the intermediate low dimensionalised data) and the output --- the flow fields cannot be decomposed nor visualized like POD or DMD{\Kair{, which can extract the individual representation of low-dimensional mapping}}.

In this study, we present a new flow decomposition method {\Kair based on a CNN autoencoder which can take account nonlinearity into its structure} in order to decompose flow fields into nonlinear low dimensional {\it modes} and to visualize each mode.
We apply this method to a flow around a circular cylinder at $\Rey_D=100$ 
to clarify what the network actually learns about the flow.

\vspace{-4mm}
\section{Methods}
\subsection{Training data}

The training data are obtained by a two-dimensional direct numerical simulation (DNS) of flow around a circular cylinder. 
The governing equations are the incompressible continuity and Navier--Stokes equations,
\begin{align}
    &\bm{\nabla} \cdot \bm{u}=0, \\
    &\dd{\bm{u}}{t} = - \bm{\nabla} \cdot (\bm{uu}) - \bm{\nabla} p + \frac{1}{{Re}_D}\nabla ^2 \bm{u},
\end{align}
where $\bm{u}$ and $p$ denote the velocity vector and pressure, respectively.
All quantities are made dimensionless by the fluid density, the free-stream velocity, and the cylinder diameter.
The Reynolds number based on the cylinder diameter is $\Rey _D=100$.
The size of the computational domain is $L_x=25.6$ and $L_y=20.0$ in the streamwise ($x$) and the transverse ($y$) directions, respectively.
The origin of coordinates is defined at the centre of the inflow boundary, and the cylinder centre is located at $(x,y)=(9,0)$.
A Cartesian grid system with the grid spacing of $\Delta x=\Delta y = 0.025$ is used.
The number of grid points is $(N_x, N_y)=(1024, 800)$.
The no-slip boundary condition on the cylinder surface is imposed using the ghost cell method of \citet{kor2017}.  

In the present study, we focus on the flow{\Kair s} around the cylinder.
{\Kair For the first case with periodic vortex shedding under a statistically steady state,} we extract a part of the computational domain, i.e., $8.2 \leq x \leq 17.8$ and $-2.4 \leq y \leq 2.4$.
Thus, the number of the grid points used for machine learning is $(N_x^*, N_y^*)=(384, 192)$.  
As the input and output attributes, the fluctuation components of streamwise velocity $u$ and transverse velocity $v$ are utilized. 
The time interval of flow field data is 0.25 corresponding to approximately 23 snapshots per a cycle with the Strouhal number equals to 0.172.
{\Kair For the second case with transient wake, the computational procedure is the same, but a larger domain is used as explained later in \S3.2.}

\subsection{Machine learning model}

\begin{figure}
  \centerline{\includegraphics[clip,width=1.0\linewidth]{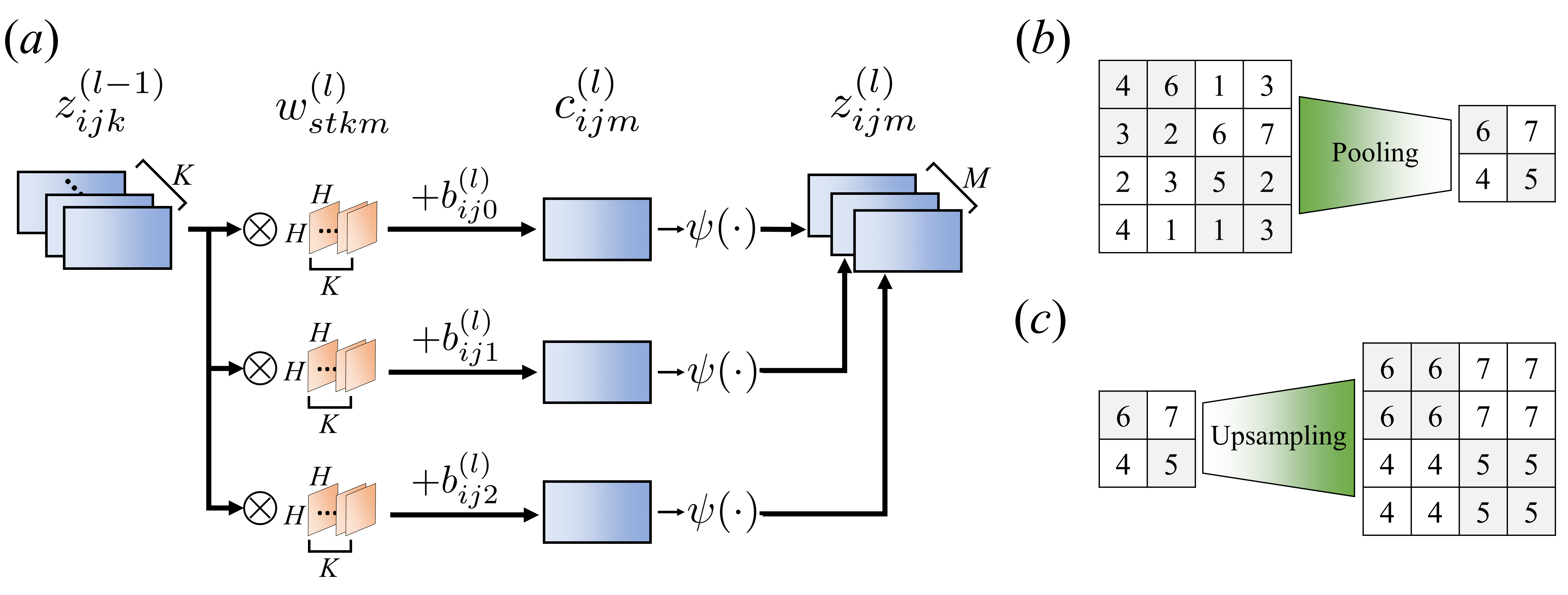}}
  \caption{Internal operations of convolutional neural network: $(a)$ convolutional layer, $(b)$ pooling layer, and $(c)$ upsampling layer.}
  \label{fig1}
\end{figure}

Convolutional neural network (CNN) mainly consists of three layers: convolutional layer, pooling layer and upsampling layer.  
The main procedure in the convolutional layer is illustrated in \figref{fig1}$(a)$.  
Using the filter with the size of $H\times H\times K$ to the input $z^{(l-1)}_{ijk}$ on the pixel represented by indices $(i,j,k)$, the filtered data $c_{ijm}$ on a pixel $(i,j,m)$ is given by
\begin{align}
    c_{ijm}^{(l)}=\sum_{k=0}^{K-1} \sum_{s=0}^{H-1} \sum_{t=0}^{H-1}
    z^{(l-1)}_{i+s,j+t,k}w_{stkm}^{(l)}+b_{ijm}^{(l)},
\end{align}
where $w_{stkm}^{(l)}$ and $b_{ijm}^{(l)}$ denote the weight and the bias at layer $l$, respectively.  In the present paper, the input and output of autoencoder model are represented as $z^{(0)}=z^{(l_{\rm max})}={\bm q}=\{u,v\}$.  
For this value, the activation function $\psi$ is applied to obtain the output of this layer:
\begin{align}
    z^{(l)}_{ijm} = {\psi}({c^{(l)}_{ijm}}).
\end{align}
In general, a nonlinear function is used as the activation function of hidden layers, as explained later.  
With the pooling layer shown in \figref{fig1}$(b)$, the data are compressed by $(1/P)^2$ times in such a way that the maximum value represents the region with the size of $P\times P$, i.e., max pooling.   
By combining the convolutional
and pooling layers, it is possible to reduce the dimension
while retaining the features of the input data.  
In the process enlarging the data dimension, the upsampling layer is used to expand the data by copying as shown in \figref{fig1}$(c)$.

\begin{figure}
  \centerline{\includegraphics[clip,width=1.0\linewidth]{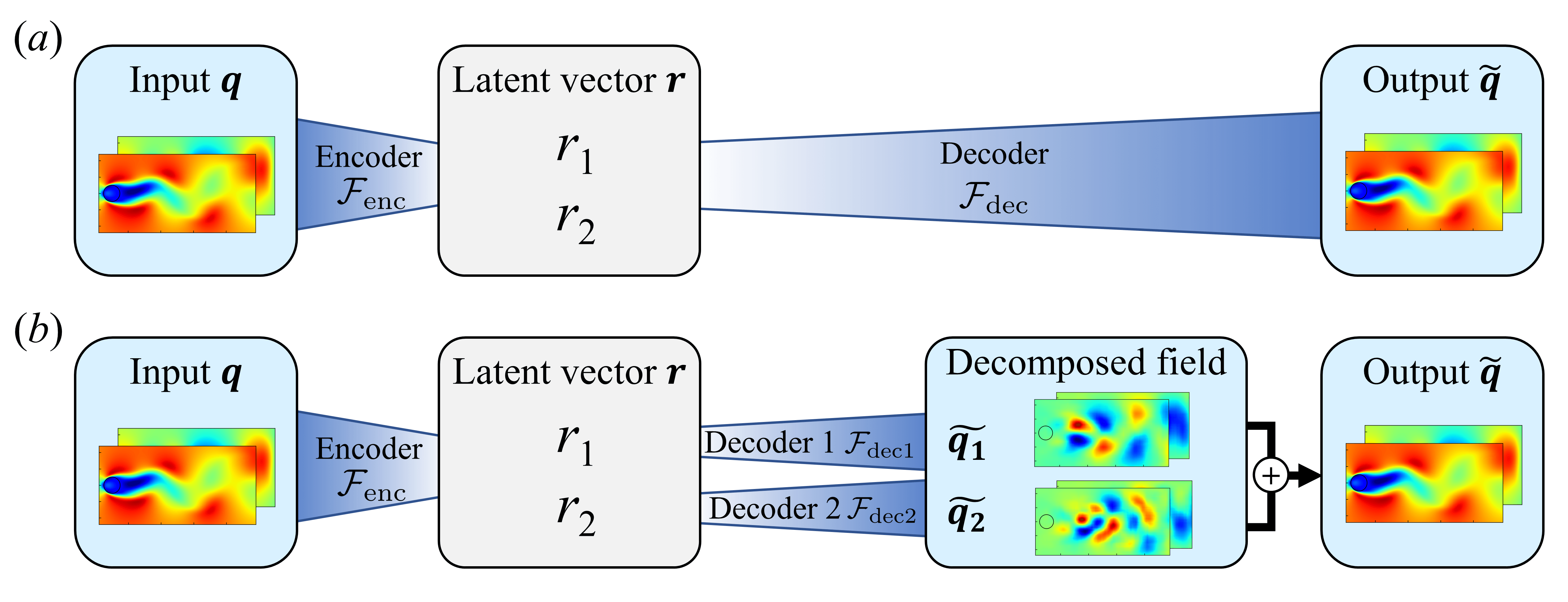}}
  \caption{Schematic of two types of CNN autoencoder used in the present study; $(a)$ conventional type CNN  {\color{black}autoencoder (C-CNN-AE)}, and $(b)$ mode decomposing CNN {\color{black}autoencoder (MD-CNN-AE)}.}
  \label{fig2}
\end{figure}

The concept of the conventional type CNN autoencoder {\color{black}(C-CNN-AE)} is illustrated in figure \ref{fig2}$(a)$. 
It consists of two parts: an encoder ${\mathcal F}_{\rm enc}$ and a decoder ${\mathcal F}_{\rm dec}$.  
The encoder works to map the high dimensional flow field into a low dimensional space. 
In the present study, we map the flow around a cylinder into two-dimensional latent space (shown as $r_1$ and $r_2$ in figure \ref{fig2}).
The decoder is used to expand the dimension from the latent space.  
In the encoder ${\mathcal F}_{\rm enc}$, the input data $\bm{q}$ with the size of $(N_x^*, N_y^*, N_\phi)=(384,192,2)$, where $N_\phi$ is the size of the feature vector, is mapped to the latent vector $\bm{r}$ with the size of $(2,1,1)$, i.e., two values.
In the decoder ${\mathcal F}_{\rm dec}$, the output data $\tilde{\bm{q}}$ having the same dimension as $\bm{q}$ is restored from the latent vector $\bm{r}$. 
Summarizing in the formula, 
\begin{align}
    \bm{r} = {\mathcal F}_{\rm enc}(\bm{q}),~~~\tilde{\bm{q}} = {\mathcal F}_{\rm dec}(\bm{r}).
\end{align}
The objective of the autoencoder is to seek the optimized weights $\bm w$ so as to minimize the $L_2$ error norm between the input and the output: $\bm{w} = {\rm argmin}_{\bm{w}} || \tilde{\bm{q}} - \bm{q} ||_2^2.$
If the original data $\bm{q}$ are successfully restored from $\bm{r}$, it suggests that the data are well represented in the dimensions of $\bm{r}$.

\begin{table*} 
    \caption{The network structure of {\color{black}MD-CNN-AE} constructed by encoder and two decoders.  Decoder 2 has the same structure as Decoder 1.}
    \begin{center}
    \begin{tabular}{cccc}
        \hline
        \multicolumn{2}{c}{Encoder} & \multicolumn{2}{c}{Decoder 1} \\ \hline
        Layer & Data size  & Layer & Data size \\ \hline
        Input & $(384,192,2)$ & 1st Value & $(1,1,1)$  \\
        1st Conv. $(3,3,16)$ & $(384,192,16)$ & Fully-connected & $(6,3,4)$ \\
        1st MaxPooling & $(192,96,16)$ & 1st Upsampling & $(12,6,4)$\\
        2nd Conv. $(3,3,8)$ & $(192,96,8)$ & 7th Conv. $(3,3,4)$ & $(12,6,4)$\\
        2nd MaxPooling & $(96,48,8)$ & 2nd Upsampling & $(24,12,4)$\\
        3rd Conv. $(3,3,8)$ & $(96,48,8)$ & 8th Conv. $(3,3,8)$ & $(24,12,8)$\\
        3rd MaxPooling & $(48,24,8)$ & 3rd Upsampling & $(48,24,8)$\\
        4th Conv. $(3,3,8)$ & $(48,24,8)$ & 9th Conv. $(3,3,8)$ & $(48,24,8)$\\
        4th MaxPooling & $(24,12,8)$ & 4th Upsampling & $(96,48,8)$ \\
        5th Conv. $(3,3,4)$ & $(24,12,4)$ & 10th Conv. $(3,3,8)$ & $(96,48,8)$\\
        5th MaxPooling & $(12,6,4)$ & 5th Upsampling & $(192,96,8)$\\
        6th Conv. $(3,3,4)$ & $(12,6,4)$ & 11th Conv. $(3,3,16)$ & $(192,96,16)$\\
        6th MaxPooling & $(6,3,4)$ & 6th Upsampling & $(384,192,16)$\\
        Fully-connected & \multirow{2}{*}{$(2,1,1)$} & 12th Conv. $(3,3,2)$ & \multirow{2}{*}{$(384,192,2)$}\\
        (Latent vector) &  & (Decomposed field 1) &  \\
        \hline
    \end{tabular}
    \end{center} \label{tab1}
    \vspace{-3mm}
\end{table*}
\begin{table*}
\centering
\caption{{\Kair Hyper parameters used for the present MD-CNN-AE.}}
  \begin{tabular}{cccc} \hline
    Parameter & Value & Parameter & Value \\ \hline
    CNN filter size & $3 \times 3$ &
    Batch size & 100 \\
    CNN pooling size & $2 \times 2$ &
    Optimizer for network & Adam \\
    Number of layers & 28 &
    Learning rate of Adam & 0.001 \\
    Number of training data & 10000 &
    $\beta_1$ of Adam & 0.9 \\
    Time interval of data & 0.25 &
    $\beta_2$ of Adam & 0.999 \\
    Percentage of training data & 70\% &
    Learning rate decay of Adam & 0 \\
    Number of epochs & 2000 & & \\
    \hline
  \end{tabular}
  \label{tab:param}
\end{table*}


In the {\Kair C-CNN-AE}, the dimension reduction of data can be done, but the intermediate output data are hard to interpret because the weights are randomly optimized during the process of training.  
Thus, we propose a {\it mode decomposing} CNN autoencoder {\color{black}(MD-CNN-AE)} shown in figure \ref{fig2}$(b)$.
The encoder part of {\color{black}MD-CNN-AE} is similar to that of {\Kair C-CNN-AE}, but the latent vector $\bm{r}$ is divided into two variables, $r_1=r_{1,1,1}$ and $r_2=r_{2,1,1}$, where the subscripts denote the indices of $\bm{r}$.  
The first decoder ${\mathcal F}_{\rm dec1}$ is used to make the first decomposed field $\tilde{\bm{q}_1}$ from the first variable $r_1$ and
the same for the second decoder ${\mathcal F}_{\rm dec2}$, i.e., $\tilde{\bm{q}_2}$ from $r_2$.
The summation of two decomposed fields, $\tilde{\bm{q}_1}$ and $\tilde{\bm{q}_2}$, is the output $\tilde{\bm{q}}$ of {\color{black}MD-CNN-AE}. 
In sum, the process are 
\begin{align}
    \bm{r} &={\mathcal F}_{\rm enc}(\bm{q}),\\
    \tilde{\bm{q}_1} &={\mathcal F}_{\rm dec1}(r_1), \\
    \tilde{\bm{q}_2} &={\mathcal F}_{\rm dec2}(r_2), \\
    \tilde{\bm{q}} &=\tilde{\bm{q}_1}+\tilde{\bm{q}_2}.
\end{align}
Since {\color{black}MD-CNN-AE} has the same structure as POD in the sense that it obtains the fields for each low dimensional mode and adds them, it can decompose flow fields in such a way that each mode can be visualized, which cannot be done with {\Kair C-CNN-AE}.

As the network parameters mentioned above, we choose the filter length $H=3$ and $K=2$ corresponding to ${\bm q}=\{u,v\}$, the max pooling ratio $P=2$, and the number of the layers $l_{\rm max}=28$.
The details of the proposed machine learning models are summarized in table \ref{tab1}.
{\Kair The number of trainable parameters for the present MD-CNN-AE is 9646.}
For training both CNNs, we appl{\Kair y} the early stopping criterion \citep{prechelt1998} to avoid overfitting and use the Adam algorithm \citep{kingma2014} to seek the optimized weights $\bm w$.
In the training process, randomly chosen 7000 snapshots of data {\Kair are} used as training data and 3000 snapshots were used as validation data. {\color{black}Five-fold cross validation \citep{BruntonKutz2019} is performed for making all machine learning models in the present study, although only the results of a single case will be shown for brevity.}
{\Kair The other hyper parameters used in the present study are summarized in table~\ref{tab:param}.}
{\Kair For further details on the implementation of {\color{black}MD-CNN-AE}, interested readers are referred to the sample Python code available on our project webpage (http://kflab.jp/en/index.php?18H03758).}  

\vspace{-2mm}
\section{Results and discussion}
\subsection{\Kair Periodic vortex shedding case}

\begin{figure}
  \centerline{\includegraphics[clip,width=1.00\linewidth]{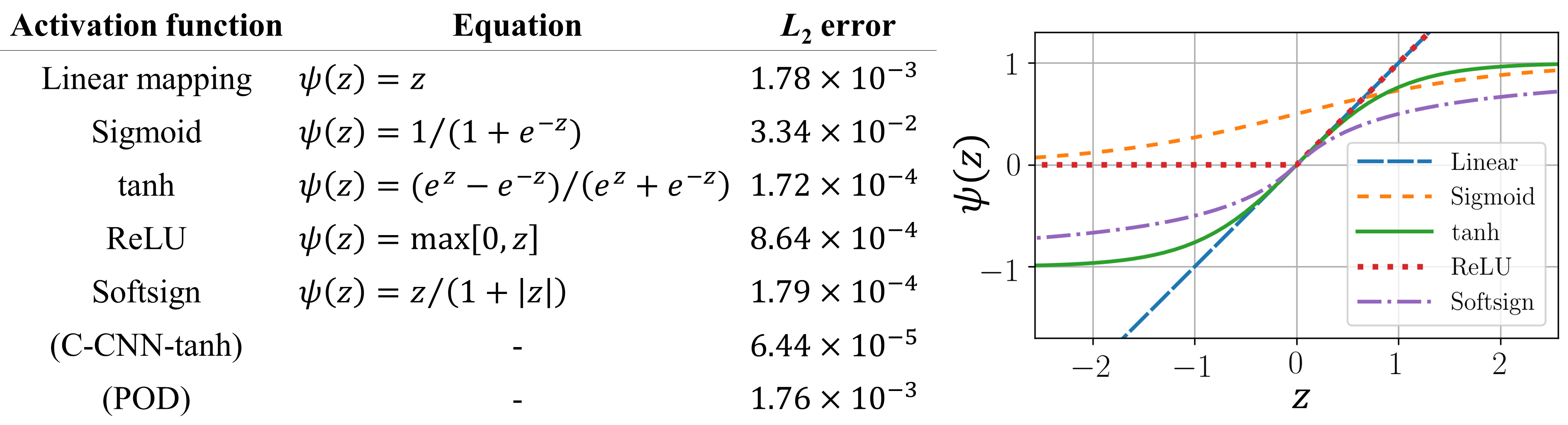}}
  \caption{Activation functions used in the present study and $L_2$ norm error for each method.}
  \label{fig3}
\end{figure}

First, we examine the {\color{black}MD-CNN-AEs} with different activation functions: linear activation, rectified linear unit (ReLU), hyperbolic tangent function (tanh), standard sigmoid function (Sigmoid), and softsign function (Softsign), as summarized in figure \ref{fig3}.  
In this figure, we also present the $L_2$ norm errors calculated by 2000 test snapshots in these five {\color{black}MD-CNN-AEs}, excluding the training process, and compare them with the cases of {\Kair C-CNN-AE} with tanh activation (C-CNN-tanh) and POD with the first two modes only.  
In the case of Sigmoid, the network is not trained well because of the vanishing gradient problem.  
The CNN with linear activation has the same error level as POD, which suggests that the linear CNN is also similar to POD as is in the case of fully connected multi-layer perceptrons with linear activation \citep{BH1989,Milano2002}.
When the nonlinear activation function (ReLU, tanh or Softsign) is used, the errors are less than that of linear activation and POD.  
Among them, tanh and Softsign, which have higher nonlinearity, result in lower $L_2$ norm errors.  
From these results, it is confirmed that the nonlinearity is the key to improve the performance of the model. 
Comparing the network structures under the same activation function (i.e., tanh), {\color{black}MD-CNN-tanh} has a slightly larger error than {\color{black}C-CNN-tanh} because of its complex structure.
In the following, we compare the results obtained by {\Kair MD-CNN-AE} with linear activation (MD-CNN-Linear) and that with tanh (MD-CNN-tanh) to investigate the effect of nonlinearity.
{\color{black} Note that the similar trends are observed in the other iterations for cross validation.}

\begin{figure}
  \centerline{\includegraphics[clip,width=1.0\linewidth]{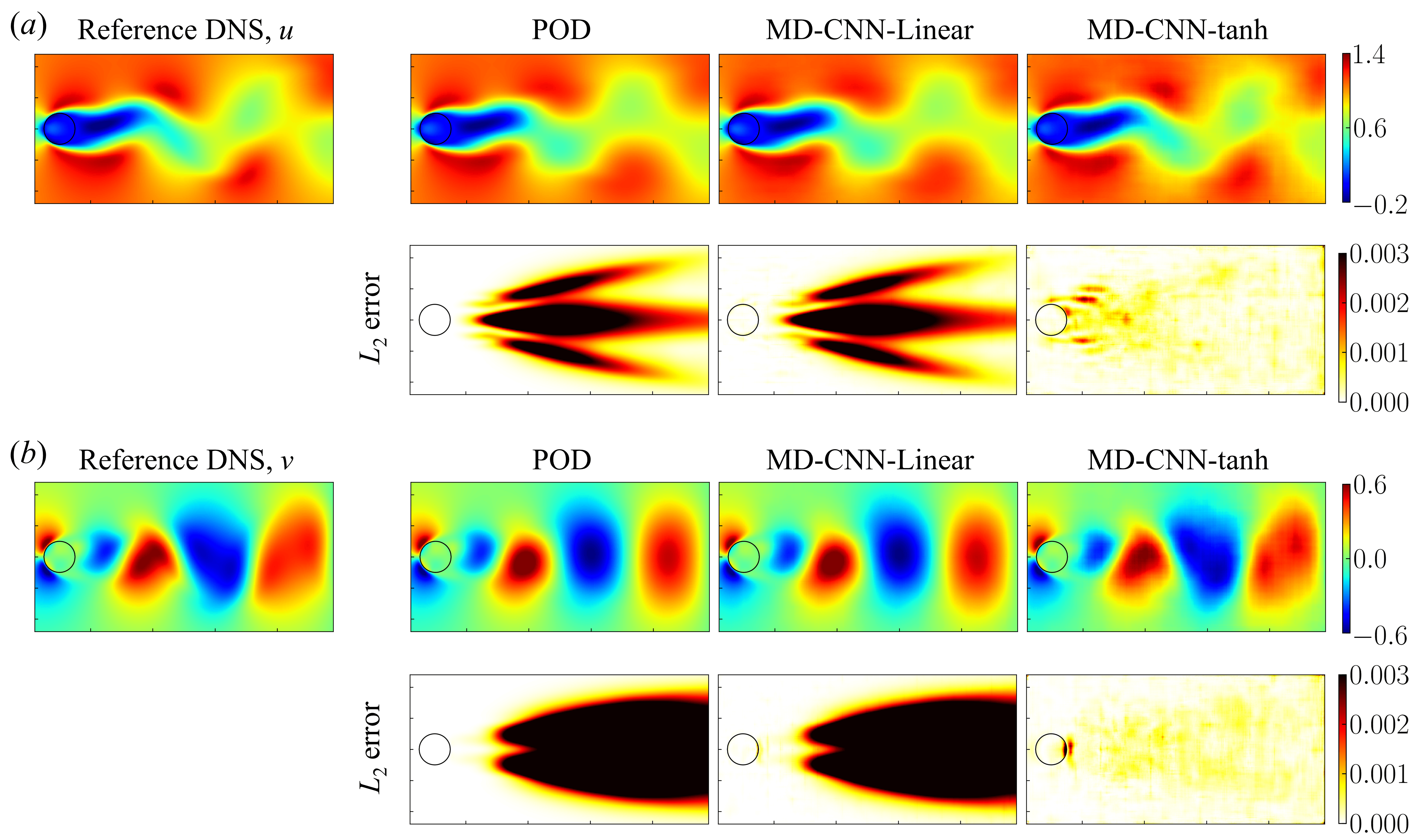}}
  \caption{The reference instantaneous flow field, output flow field and distribution of $L_2$ norm error in three methods: $(a)$ streamwise velocity $u$, $(b)$ transverse velocity $v$.}
  \label{fig4}
\end{figure}

The output of the machine-learned models (MD-CNN-Linear and MD-CNN-tanh) and POD are summarized in figure \ref{fig4}. 
The flow fields reconstructed by all three methods show reasonable agreements with the reference data.  
The field reconstructed by MD-CNN-tanh is closest to the reference.
Interestingly, the reconstructed fields of MD-CNN-Linear and POD are similar, which confirms the similarity mentioned above.

In order to evaluate the reconstruction error, we assess the time-averaged local $L_2$ norm error with 2000 test snapshots excluding the training process, as shown in figure \ref{fig4}. 
Comparing three methods, MD-CNN-tanh shows the lowest error in the entire region except for the very small region downstream of the cylinder.  
The distributions of $L_2$ norm error in POD and MD-CNN-Linear are again similar due to their similarity mentioned above.  

\begin{figure}
  \centerline{\includegraphics[clip,width=1.0\linewidth]{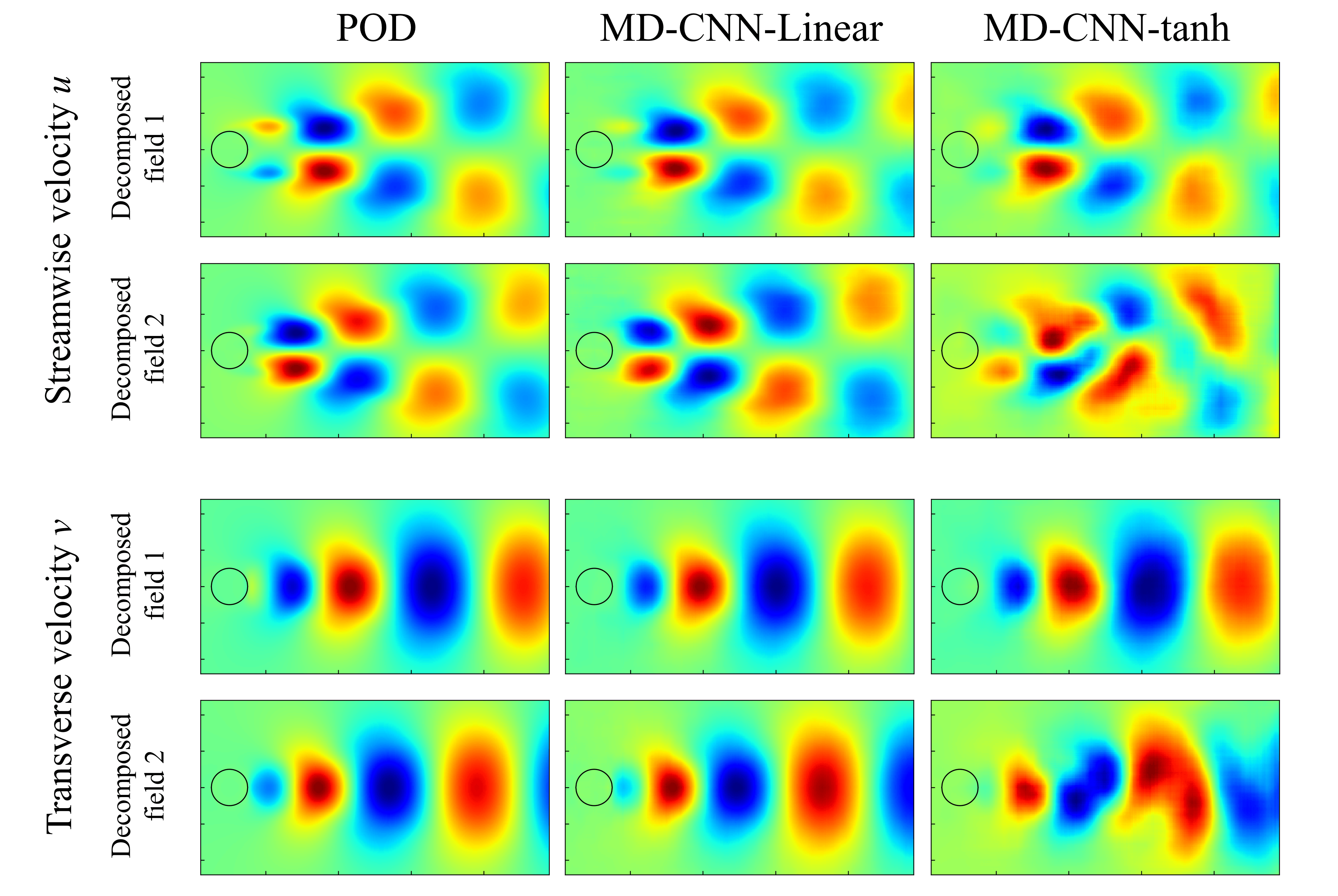}}
  \vspace{-2mm}
  \caption{The decomposed flow fields with POD, MD-CNN-Linear, and MD-CNN-tanh.}
  \label{fig5}
\end{figure}

\begin{figure}
  \centerline{\includegraphics[clip,width=0.7\linewidth]{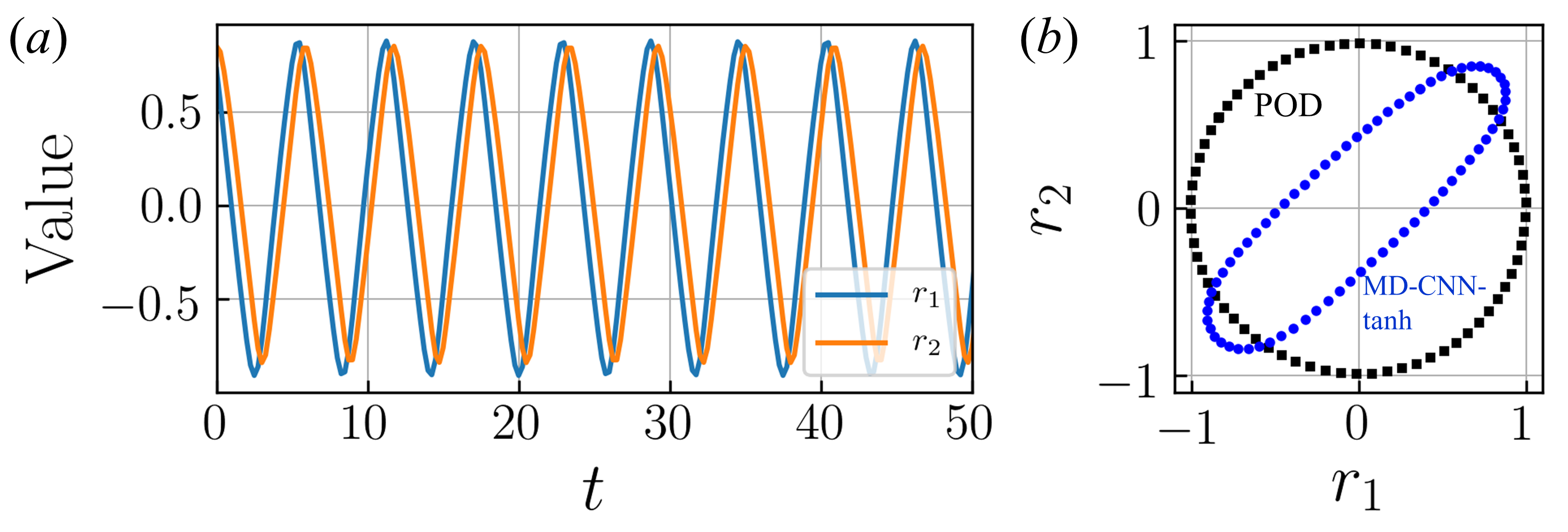}}
  \caption{{\color{black} Encoded variables ($r_1, r_2$) with MD-CNN-tanh: $(a)$ time traces, $(b)$ trajectory compared with that of POD.}}
  \label{fig6}
\end{figure}

The strength of the present {\color{black}MD-CNN-AE} over the conventional CNN is that the flow field can be decomposed and visualized.
Figure \ref{fig5} visualizes the two decomposed fields corresponding to the velocity distributions of figure \ref{fig4}.  
Note that the time-averaged component of decomposed fields is subtracted in {\color{black}MD-CNN-AEs}.  
The decomposed field of POD and that of MD-CNN-Linear are almost the same, and the decomposed field of MD-CNN-tanh is distorted, likely due to the nonlinearity of the activation function.
{\Kair{
Figure~\ref{fig6} shows the time traces of the corresponding encoded variables ($r_1, r_2$) by MD-CNN-tanh and compares the trajectory with that of POD.
The encoded variables obtained by MD-CNN-tanh are also periodic in time, corresponding to the vortex shedding, but the phases are observed to be shifted from that of POD.
It is worth noting that, although not shown here, the periodic signals of $r_1$ and $r_2$ are observed to be similar in the five-fold cross validation but the amount of phase shift (i.e., trajectory) is not unique.
It suggests that the decomposition by MD-CNN-tanh is not unique due to the nonlinearity.
}}

\begin{figure}
  \centerline{\includegraphics[clip,width=\linewidth]{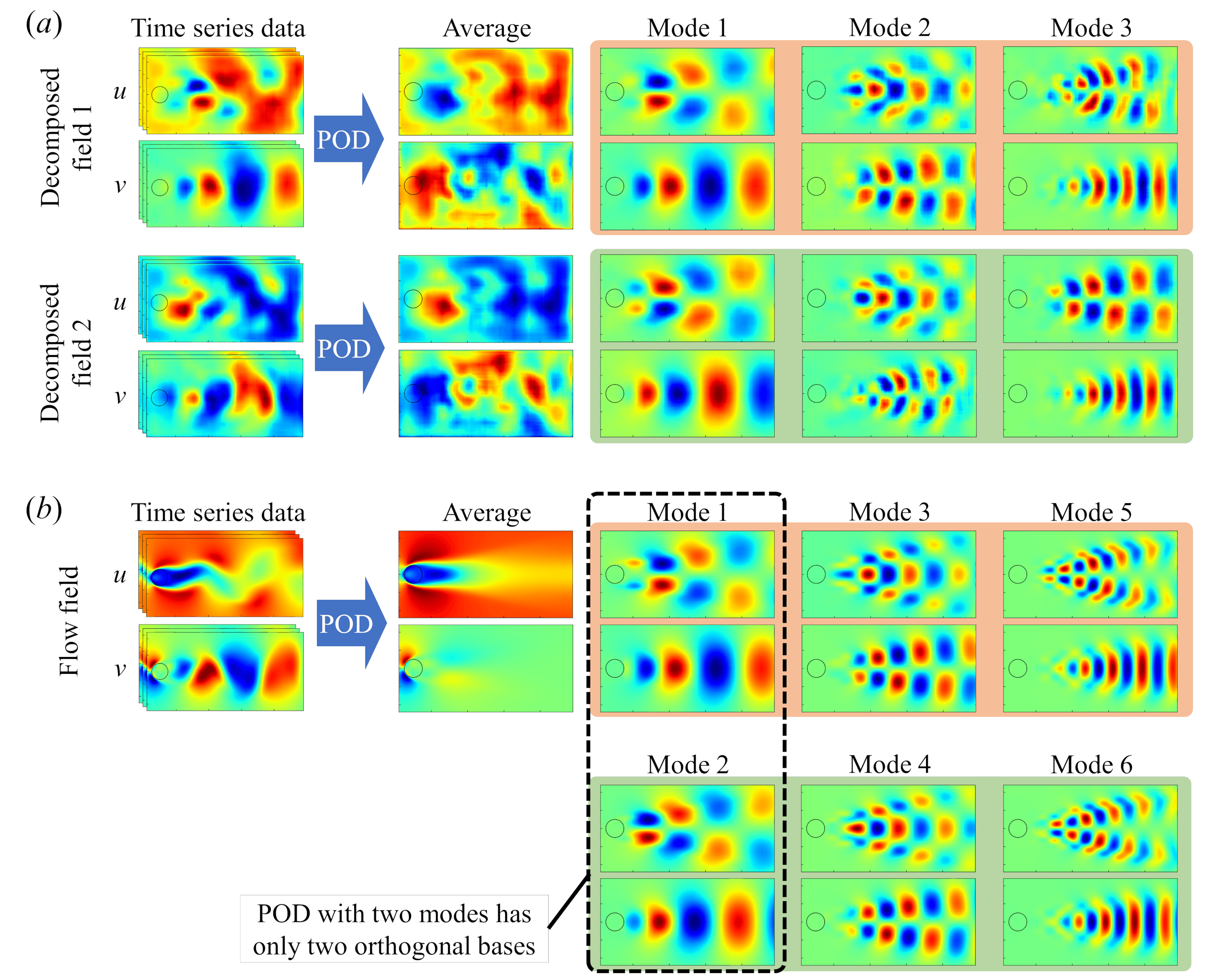}}
  \caption{The POD orthogonal basis of two decomposed fields of MD-CNN-tanh and reference DNS.}
  \label{fig7}
\end{figure}

To examine {\Kair the distortion observed above} further, we perform POD for the decomposed fields obtained by the MD-CNN-tanh model, as shown in figure \ref{fig7}$(a)$.  
We also present in figure \ref{fig7}$(b)$ the POD results of reference flow field to compare with the machine-learned model.   
The interesting view is that decomposed field 1 contains the orthogonal bases akin to POD modes 1, 3 and 5, and decomposed field 2 contains modes 2, 4 and 6.
Note that complicated structures observed in the average fields shown in figure \ref{fig7}$(a)$ are mostly canceled out by adding these decomposed fields.
It suggests that the proposed method also decomposes the average field of the fluctuation components which should be zero via nonlinear function.  
It is also worth noting that the ratio of the amounts of kinetic energy contained in decomposed field 1 and decomposed field 2 are nearly equal.

\begin{figure}
  \centerline{\includegraphics[clip,width=1.0\linewidth]{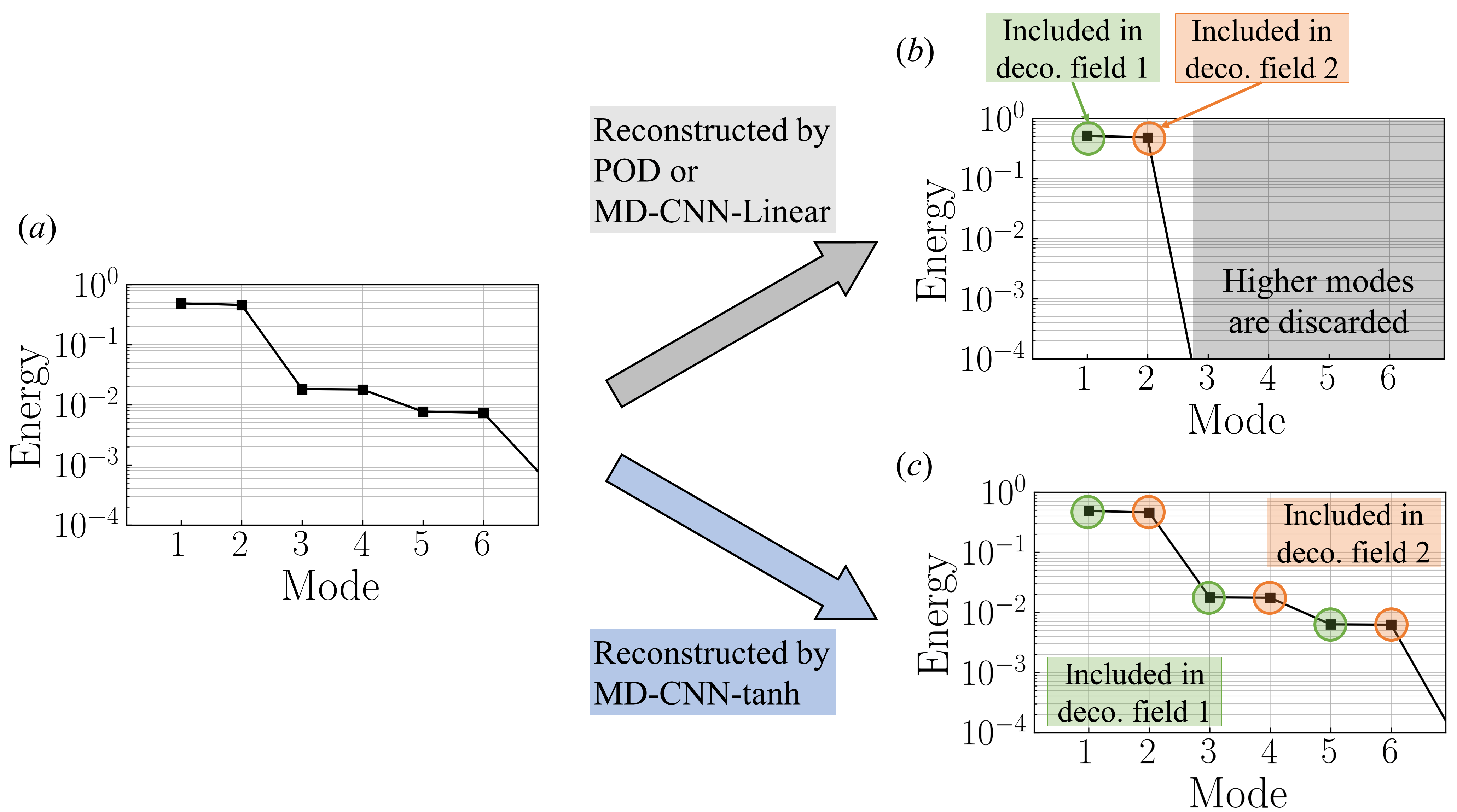}}
  \caption{Normalized value of the energy distribution of the orthogonal basis of $(a)$ flow field, $(b)$ reconstructed field using POD with two modes only or MD-CNN-Linear, and $(c)$ reconstructed field using MD-CNN-tanh.}
  \label{fig8}
\end{figure}

Let us present in figure \ref{fig8} the normalized values of the energy distribution of the orthogonal bases contained in the flow field.  
When we use only the first two POD modes to reconstruct the flow field --- of the matter of course, though --- decomposed field 1 consists of mode 1, and decomposed field 2 consists of mode 2, while higher modes are discarded as indicated by the gray area of figure \ref{fig8}$(b)$. 
The situation is the same for MD-CNN-Linear.
On the other hand, for MD-CNN-tanh, the two decomposed fields contain multiple POD modes, and the characteristics of higher modes are retained, which results in the lower reconstruction error than the POD with first two modes only. 
In addition, the flow field is decomposed in such a way that the orthogonal bases are distributed to two decomposed fields in a similar manner as the full POD, as shown in figure \ref{fig8}$(c)$.

In the present example problem of two-dimensional flow around a cylinder cylinder, it is known that the third to sixth POD modes can be expressed by analytical nonlinear functions of the first two POD modes \citep{Loiseau2018}.
The present result with MD-CNN-tanh is consistent with this knowledge, and it suggests that such nonlinear functions are embedded in the nonlinearity of MD-CNN-tanh.

{\Kair

\begin{figure}
  \centerline{\includegraphics[clip,width=1.0\linewidth]{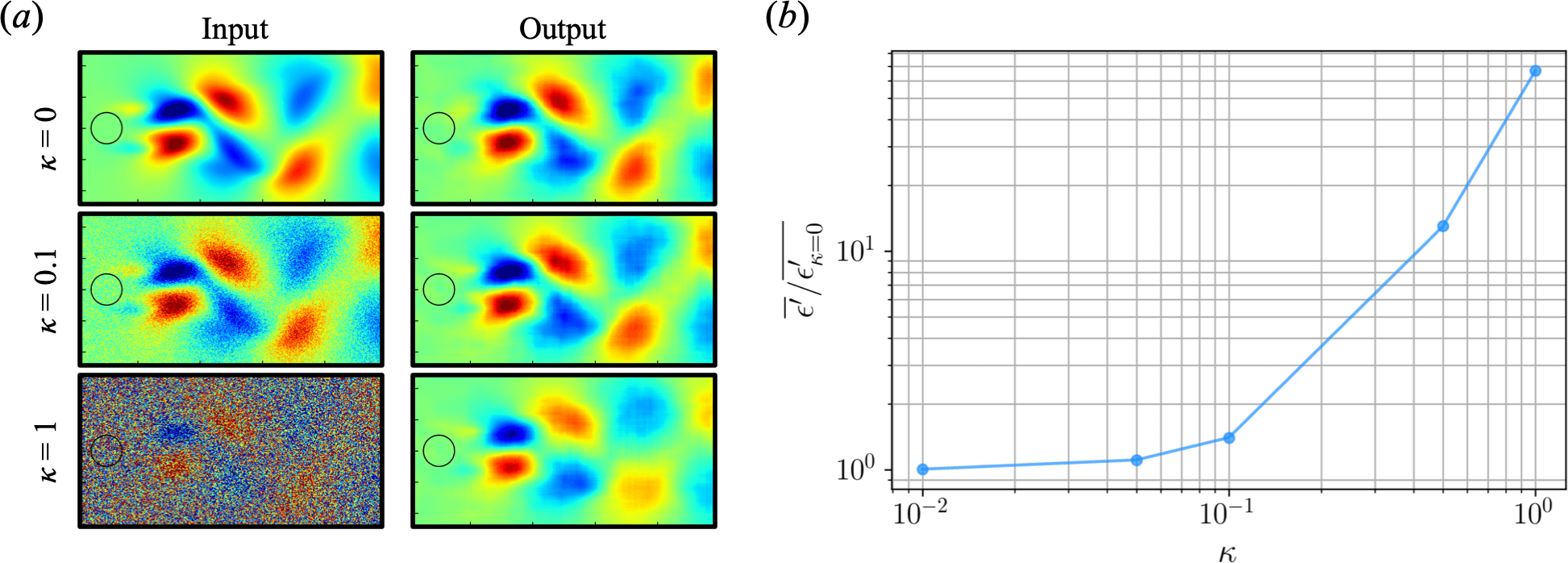}}
  \caption{\Kair{Demonstration of the robustness analysis for noisy inputs with MD-CNN-tanh. $(a)$ Streamwise velocity fluctuation $u^\prime$ with $\kappa=0$ (without noise), $0.1$, and $1$.  $(b)$ $\kappa$-$(\overline{\epsilon^\prime}/\overline{\epsilon^\prime_{\kappa=0}})$ plot.  The five-fold cross validation is made although not shown here.}}
  \label{fig9}
\end{figure}

We also examine the robustness of the MD-CNN-tanh for a noisy input in order to assess the applicability to experimental situations, as shown in figure \ref{fig9}.  
Here, let the $L_2$ norm error for a noisy input be $\epsilon^{\prime}=||{\bm q}^{\prime}_{\rm DNS}-{\cal F}({\bm q}^{\prime}_{\rm DNS}+\kappa {\bm n})||_2^2$, where $\bm q$ is the feature vector, ${\bm n}$ is the Gaussian random noise with unit variance, and $\kappa$ is a magnitude of noisy inputs.  With $\kappa=0.1$, the output of MD-CNN-tanh shows reasonable agreement with the input DNS data.  
Over $\kappa=0.1$, the error drastically increases with increasing the magnitude of noise $\kappa$, as shown in figure \ref{fig9}$(b)$.  
Noteworthy here is that the MD-CNN-tanh has a denoising effect as observed for $\kappa=0.1$ and $1$ of figure \ref{fig9}$(a)$.  
A similar observation has been reported by \citet{erichson2019}, who applied multi-layer perceptrons to a cylinder wake.
}

\begin{figure}
  \centerline{\includegraphics[clip,width=0.8\linewidth]{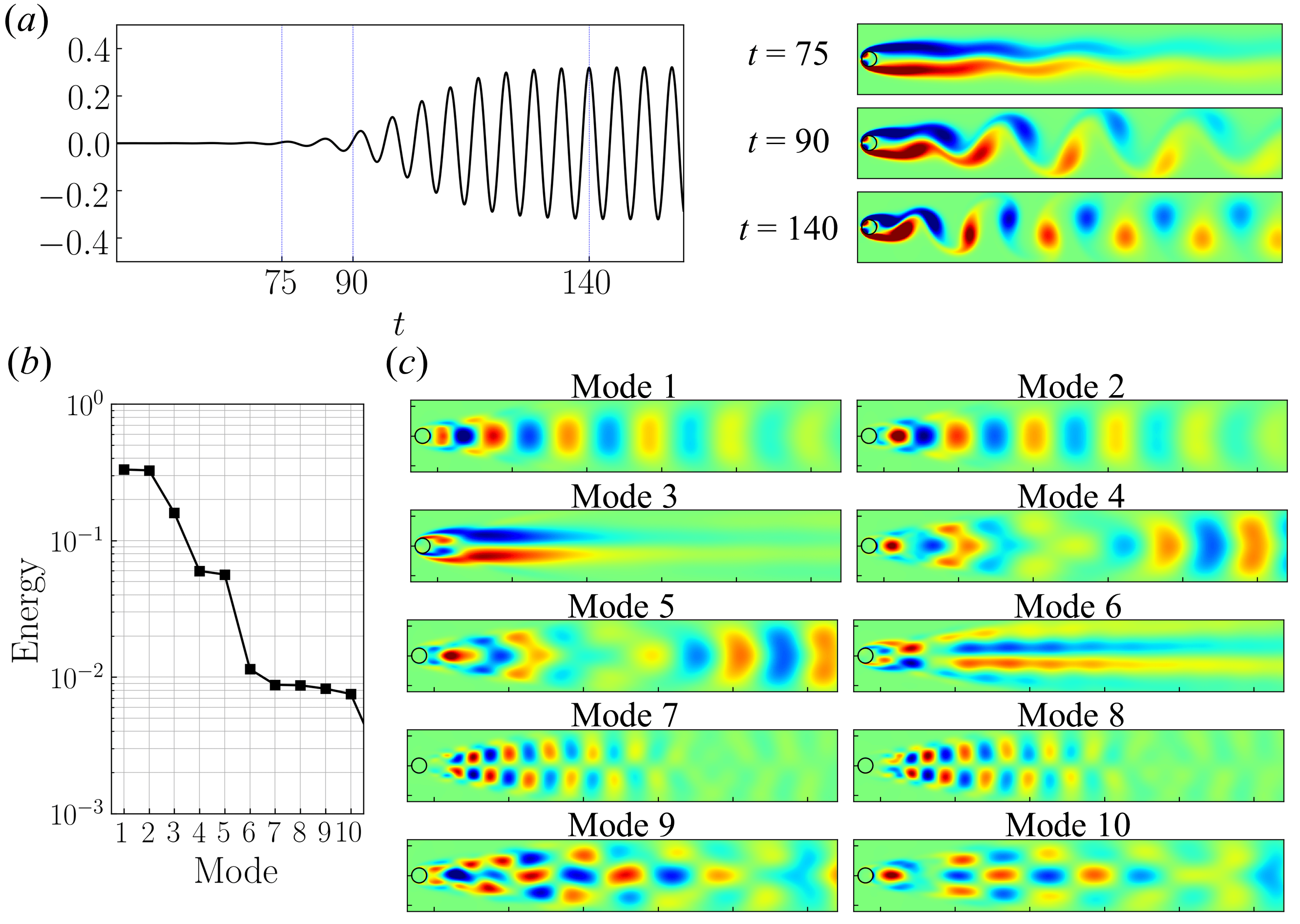}}
  \caption{\color{black}($a$) The lift coefficient $C_L$ of the transient process, ($b$) normalized value of the energy distribution of first ten POD modes, ($c$) the corresponding vorticity fields.}
  \label{fig10}
\end{figure}

\subsection{\Kair Transient wake case}

{\Kair
As an example of more complex flows comprising high-order modes, let us consider a transient process with a circular cylinder wake at $Re_D=100$.  
For the transient flow, the streamwise length of computational domain and that of flow field data are extended to $L_x=51.2$ and $8.2 \leq x \leq 37$, i.e., $N_x^*=1152$.
{\color{black} To focus on the transient process, we use the flow field data of $50 \leq t \leq 150$ with the time step of $\Delta t=0.025$.}
The temporal development of the lift coefficient $C_L$ and the energy and vorticity fields of the first ten POD modes are summarized in figure \ref{fig10}.
All of these quantities exhibit the trends similar to those of \citet{NSMS2016}.

\begin{figure}
  \centerline{\includegraphics[clip,width=1.0\linewidth]{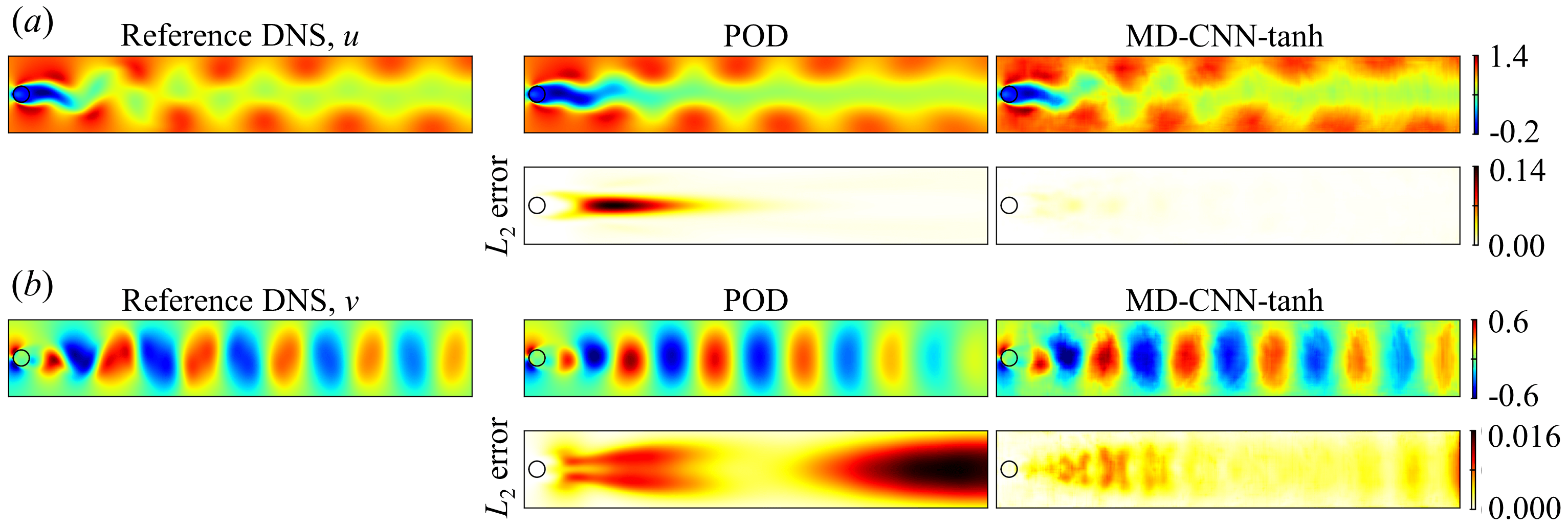}}
  \caption{\color{black} The reference instantaneous flow field, output flow field and distribution of $L_2$ norm error at $t=137.5$ with two methods: ($a$) streamwise velocity $u$, and ($b$) transverse velocity $v$.}
  \label{fig11}
\end{figure}

Figure \ref{fig11} compares the reference DNS flow fields and the fields reconstructed by POD and MD-CNN-tanh using the first two modes.
Similarly to the results shown in figure \ref{fig4}, the proposed method shows lower $L_2$ error than POD.

\begin{figure}
  \centerline{\includegraphics[clip,width=1.0\linewidth]{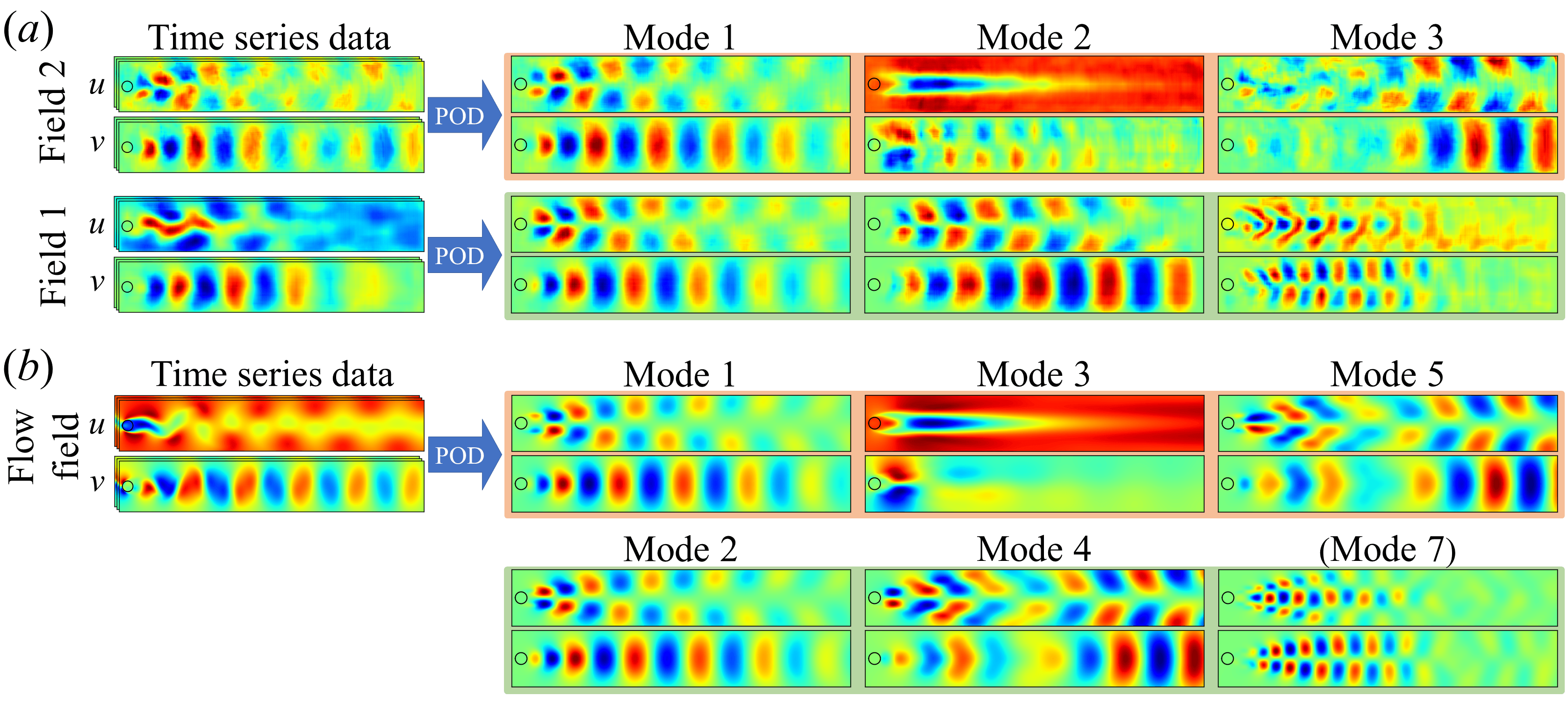}}
  \caption{\color{black} The POD orthogonal basis of ($a$) two decomposed fields of MD-CNN-tanh and ($b$) reference DNS in a transient flow.}
  \label{fig12}
\end{figure}

\begin{figure}
  \centerline{\includegraphics[clip,width=0.6\linewidth]{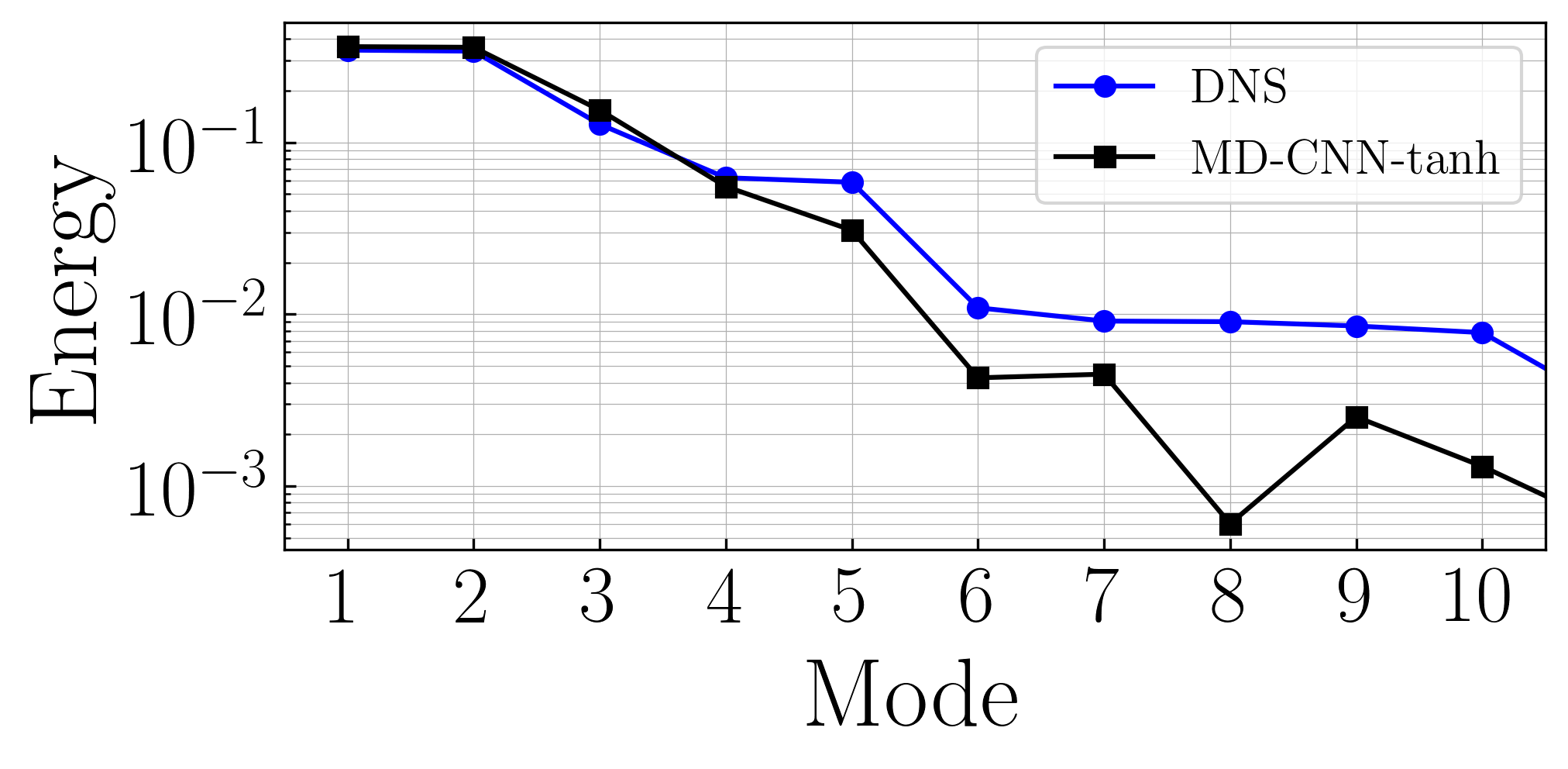}}
  \caption{\color{black} The energy distribution of the DNS and output field with MD-CNN-tanh.}
  \label{fig13}
\end{figure}

In figure \ref{fig12}, we summarize the results of performing POD to the decomposed field 1 and field 2 obtained by MD-CNN-tanh, compared with the POD modes obtained from DNS data, like figure~\ref{fig7}.  
Here, the average fields of time series data are omitted and the decomposed field 2 is shown on the top for clarity of illustration.  
The energy distribution of the output field of MD-CNN-tanh with two latent vectors, obtained by POD, is shown in figure \ref{fig13}.
Similarly to figure \ref{fig7}$(b)$, one decomposed field (field 2 in this figure) contains the orthogonal bases like POD mode 1, 3 and 5, and another decomposed field contains POD mode 2, 4 and 7.  
The difference from figure \ref{fig7}$(b)$ is that a mode resembling POD mode 7 appears in this case instead of mode 6.  
This model likely estimates that mode 7 has higher energy than mode 6, since POD mode 7 has the same degree of energy as mode 6 as shown in figure \ref{fig13}.
In sum, in this transient case, too, a single nonlinear mode of MD-CNN-tanh contains multiple POD modes in a manner broadly similar to the periodic vortex shedding case, although the correspondence to the POD modes is slightly less clear.

\begin{figure}
  \centerline{\includegraphics[clip,width=0.6\linewidth]{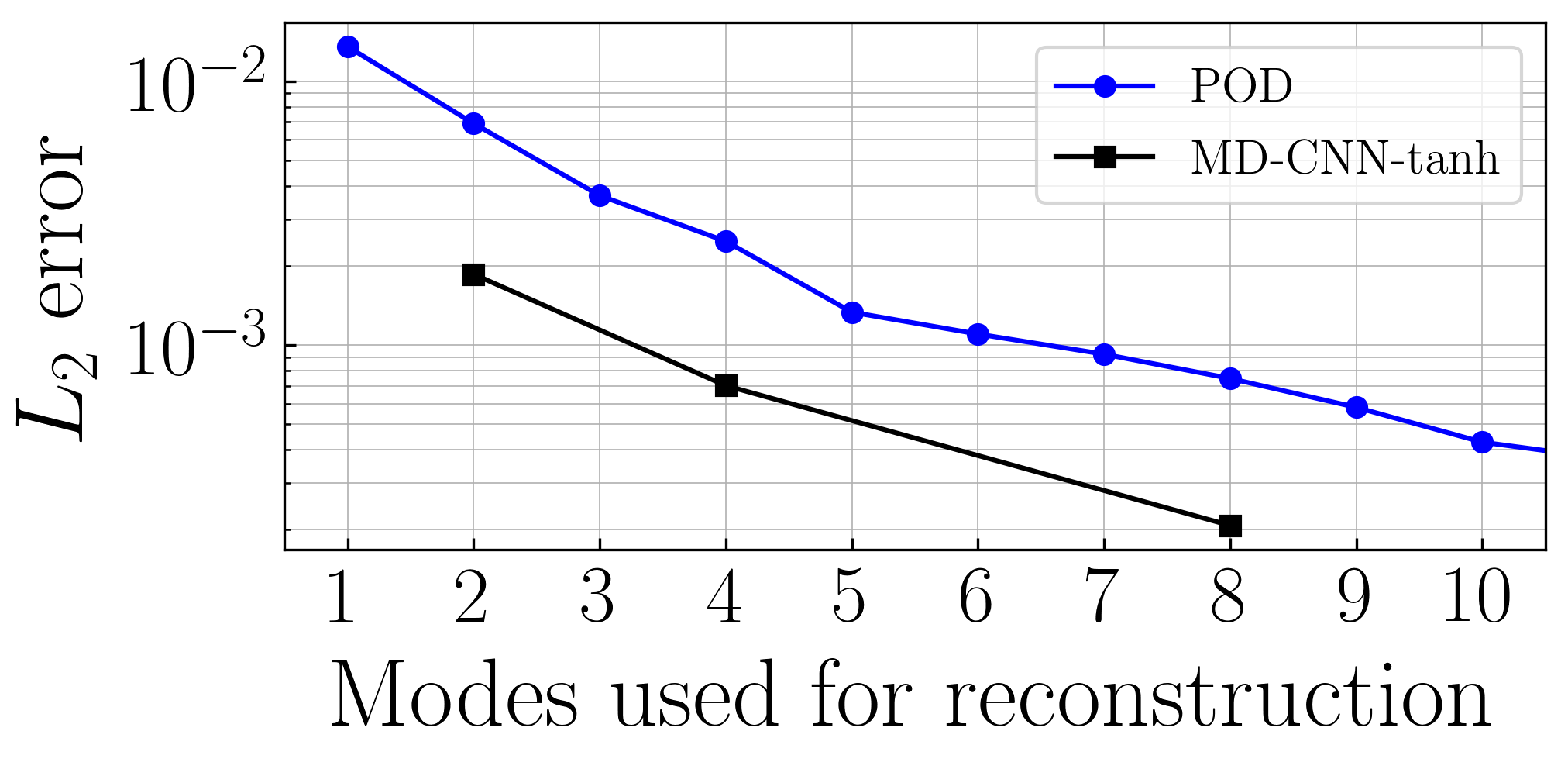}}
  \caption{{\color{black} Dependence on the number of latent vector (i.e., mode used for reconstruction) for the transient wake problem.  The five-fold cross validation is made although the error bars are not shown here.}}
  \label{fig14}
\end{figure}

{Finally, as a preliminary investigation toward extention of the present method, we show in figure \ref{fig14} a result on the dependence on the number of latent vector $n_L$ for the transient wake problem, although further investigation for $n_L>2$ cases is left as future work.
Here, we compare POD and MD-CNN-AEs with $n_L=$ 2, 4, and 8.
It is observed that the $L_2$ error of MD-CNN-AE is systematically less than that of POD with the same number of modes.
This result suggests that the present model can map the high-dimensional data into lower dimensional space than POD while retaining the feature of the unsteady flow.
However, the present result also implies that the ability of MD-CNN-AE to represent the flow with fewer modes is not as good as the most advanced nonlinear dimensionality reduction method such as the locally linear embedding (LLE) \citep{Roweis2000}, by which \citet{Ehlert2019} has very recently reported that resonsruction using 2 LLE coordinates results in much less $L_2$ error than that using 10 POD modes.
} 
}

\section{Conclusions}
As a CNN structure which can decompose flow fields in a nonlinear manner and visualize the decomposed fields,
we constructed a mode decomposing CNN autoencoder {\color{black} (MD-CNN-AE)} with one encoder and two decoders. 
As a test case, the method was applied to flow around a circular cylinder at $Re _D=100$, and the flow field was mapped into two values and restored by adding the two decomposed fields.  
With MD-CNN-Linear, which has the linear activation functions, the reconstructed field is similar to that of POD with the first two modes, both in terms of $L_2$ norm error and the distribution of reconstruction error.
It suggests that the linear CNN is also similar to POD as is in the case of linear multi-layer perceptrons.
When we use the nonlinear activation function, $L_2$ norm error of reconstruction was reduced as compared to those of POD with two modes and MD-CNN-Linear.

We also investigated the decomposed fields obtained by {\color{black}MD-CNN-AE}.  
The two decomposed fields in MD-CNN-Linear are similar to those of POD with two modes.
For MD-CNN-tanh, complex structures were observed and the two decomposed fields of MD-CNN-tanh were found to have the same amount of energy. 
By performing POD to these two decomposed fields, it was revealed that decomposed field 1 contains the orthogonal bases corresponding to POD modes 1, 3 and 5, and  decomposed field 2 contains modes 2, 4 and 6.  
The present result is also consistent with the existing knowledge on the relationship between the first two POD modes and the third to sixth POD modes in the present problem ---
it suggests that MD-CNN-tanh can be used to extract modes with lower dimensions in such a way that nonlinear functions are embedded in the network. 
{\Kair The transient process was also considered as an example of more complex flow with higher order spatial modes, and broadly similar results were obtained.} 

{\Kair Through the analysis of very simple problems, i.e., an unsteady cylinder wake and its transient process, we have confirmed the basic performance of the proposed MD-CNN-AE.  
However, the proposed method has been so far  examined  only for flows with large scale spatial structures.
To handle more complex flows, e.g., turbulence, additional improvements on the network design should be required.} 
{\Kair Nevertheless, we believe} that by extending the present idea of {\color{black} MD-CNN-AE} with nonlinear function, which can represent more information against linear theory with same number of {\it modes}, we may be able to take greater advantage of machine learning for reduced-order modeling of three-dimensional unsteady and turbulent flows,
{\Kair that can eventually be utilized for development of efficient flow control laws based on nonlinear reduced-order models}.

\section*{Acknowledgements}
Authors are grateful to Dr.~S.~Obi, Dr.~K.~Ando, and Mr.~K.~Hasegawa (Keio University) for fruitful discussions,
{\Kair and Dr. K. Zhang (UCLA) for comments and advice for the analysis of transient flows.}
K.~Fukagata also thanks Dr.~K.~Taira (UCLA) for advising K.~Fukami during his exchange.
This work was supported through JSPS KAKENHI Grant Number 18H03758 by Japan Society for the Promotion of Science.

\bibliographystyle{jfm}
\bibliography{jfm}

\begin{thebibliography}{34}
\expandafter\ifx\csname natexlab\endcsname\relax\def\natexlab#1{#1}\fi
\def\au#1{#1} \def\ed#1{#1} \def\yr#1{#1}\def\at#1{#1}\def\jt#1{\textit{#1}}
  \def\bt#1{#1}\def\bvol#1{\textbf{#1}} \def\vol#1{#1} \def\pg#1{#1}
  \def\publ#1{#1}\def\arxiv#1{#1}\def\org#1{#1}\def\st#1{\textit{#1}}

\bibitem[Alfonsi \& Primavera(2007)]{alfonsi2006}
{\sc \au{Alfonsi, G.} \& \au{Primavera, L.}} \yr{2007}  \at{The structure of
  turbulent boundary layers in the wall region of plane channel flow}.
  \jt{Proc. R. Soc. A}  \bvol{463}~(2078),  \pg{593--612}.

\bibitem[Baldi \& Hornik(1989)]{BH1989}
{\sc \au{Baldi, P.} \& \au{Hornik, K.}} \yr{1989}  \at{Neural networks and
  principal component analysis: Learning from examples without local minima}.
  \jt{Neural Netw.}  \bvol{2}~(1),  \pg{53--58}.

\bibitem[Bergmann {\em et~al.\/}(2005)Bergmann, Cordier \&
  Brancher]{bergmann2005}
{\sc \au{Bergmann, M.}, \au{Cordier, L.} \& \au{Brancher, J.-P.}} \yr{2005}
  \at{Optimal rotary control of the cylinder wake using proper orthogonal
  decomposition reduced-order model}.  \jt{Phys. Fluids}  \bvol{17}~(9),
  \pg{097101}.

\bibitem[Brunton \& Kutz(2019)]{BruntonKutz2019}
{\sc \au{Brunton, S.~L.} \& \au{Kutz, J.~N.}} \yr{2019} {\em Data-driven
  science and engineering: Machine learning, dynamical systems, and control\/}.
   \publ{Cambridge University Press}.

\bibitem[Brunton \& Noack(2015)]{BruntonNoack2015}
{\sc \au{Brunton, S.~L.} \& \au{Noack, B.~R.}} \yr{2015}  \at{Closed-loop
  turbulence control: Progress and challenges}.  \jt{Appl. Mech. Rev.}
  \bvol{67},  \pg{050801}.

\bibitem[Brunton {\em et~al.\/}(2019)Brunton, Noack \&
  Koumoutsakos]{Brunton2019}
{\sc \au{Brunton, S.~L.}, \au{Noack, B.~R.} \& \au{Koumoutsakos, P.}} \yr{2019}
   \at{Machine learning for fluid mechanics}.  \jt{{\rm
  arXiv:1905.11075}\hspace{-0.3em}} .

\bibitem[Duraisamy {\em et~al.\/}(2019)Duraisamy, Iaccarino \& Xiao]{DIX2019}
{\sc \au{Duraisamy, K.}, \au{Iaccarino, G.} \& \au{Xiao, H.}} \yr{2019}
  \at{Turbulence modeling in the age of data}.  \jt{Annu. Rev. Fluid. Mech.}
  \bvol{51},  \pg{357--377}.

\bibitem[Ehlert {\em et~al.\/}(2019)Ehlert, Nayeri, Morzynski \&
  Noack]{Ehlert2019}
{\sc \au{Ehlert, A.}, \au{Nayeri, C.~N.}, \au{Morzynski, M.} \& \au{Noack,
  B.~R.}} \yr{2019}  \at{Locally linear embedding for transient cylinder
  wakes}.  \jt{{\rm arXiv:1906.07822}\hspace{-0.3em}} .

\bibitem[Erichson {\em et~al.\/}(2019)Erichson, Mathelin, Yao, Brunton, Mahoney
  \& Kutz]{erichson2019}
{\sc \au{Erichson, N.~B.}, \au{Mathelin, L.}, \au{Yao, Z.}, \au{Brunton,
  S.~L.}, \au{Mahoney, M.~W.} \& \au{Kutz, J.~N.}} \yr{2019}  \at{Shallow
  learning for fluid flow reconstruction with limited sensors and limited
  data}.  \jt{{\rm arXiv:1902.07358}\hspace{-0.3em}} .

\bibitem[Fukami {\em et~al.\/}(2019{\natexlab{{\em a\/}}})Fukami, Fukagata \&
  Taira]{Fukami2019a}
{\sc \au{Fukami, K.}, \au{Fukagata, K.} \& \au{Taira, K.}}
  \yr{2019{\natexlab{{\em a\/}}}}  \at{Super-resolution reconstruction of
  turbulent flows with machine learning}.  \jt{J. Fluid Mech.}  \bvol{870},
  \pg{106--120}.

\bibitem[Fukami {\em et~al.\/}(2019{\natexlab{{\em b\/}}})Fukami, Nabae, Kawai,
   \& Fukagata]{Fukami2019b}
{\sc \au{Fukami, K.}, \au{Nabae, Y.}, \au{Kawai, K.},  \& \au{Fukagata, K.}}
  \yr{2019{\natexlab{{\em b\/}}}}  \at{Synthetic turbulent inflow generator
  using machine learning}.  \jt{Phys. Rev. Fluids}  \bvol{4},  \pg{064603}.

\bibitem[Hinton \& Salakhutdinov(2006)]{hinton2006}
{\sc \au{Hinton, G.~E.} \& \au{Salakhutdinov, R.~R.}} \yr{2006}  \at{Reducing
  the dimensionality of data with neural networks}.  \jt{Science}
  \bvol{313}~(5786),  \pg{504--507}.

\bibitem[Kingma \& Ba(2014)]{kingma2014}
{\sc \au{Kingma, D.~P.} \& \au{Ba, J.}} \yr{2014}  \at{Adam: A method for
  stochastic optimization}.  \jt{{\rm arXiv:1412.6980}\hspace{-0.3em}} .

\bibitem[Kor {\em et~al.\/}(2017)Kor, {Badri Ghomizad} \& Fukagata]{kor2017}
{\sc \au{Kor, H.}, \au{{Badri Ghomizad}, M.} \& \au{Fukagata, K.}} \yr{2017}
  \at{A unified interpolation stencil for ghost-cell immersed boundary method
  for flow around complex geometries}.  \jt{J. Fluid Sci. Technol.}
  \bvol{12}~(1),  \pg{JFST0011}.

\bibitem[Kutz(2017)]{Kutz2017}
{\sc \au{Kutz, J.~N.}} \yr{2017}  \at{Deep learning in fluid dynamics}.  \jt{J.
  Fluid Mech.}  \bvol{814},  \pg{1--4}.

\bibitem[LeCun {\em et~al.\/}(1998)LeCun, Bottou, Bengio \& Haffner]{Lecun1998}
{\sc \au{LeCun, Y.}, \au{Bottou, L.}, \au{Bengio, Y.} \& \au{Haffner, P.}}
  \yr{1998}  \at{Gradient-based learning applied to document recognition}.
  \jt{Proc. IEEE}  \bvol{86}~(11),  \pg{2278--2324}.

\bibitem[Ling {\em et~al.\/}(2016)Ling, Kurzawski \& Templeton]{LKT2016}
{\sc \au{Ling, J.}, \au{Kurzawski, A.} \& \au{Templeton, J}} \yr{2016}
  \at{Reynolds averaged turbulence modelling using deep neural networks with
  embedded invariance}.  \jt{J. Fluid Mech.}  \bvol{807},  \pg{155--166}.

\bibitem[Loiseau {\em et~al.\/}(2018)Loiseau, Brunton \& Noack]{Loiseau2018}
{\sc \au{Loiseau, J.-Ch.}, \au{Brunton, S.~L.} \& \au{Noack, B.~R.}} \yr{2018}
  \at{From the {POD-G}alerkin method to sparse manifold models}.  \jt{{\rm
  available on ResearchGate}\hspace{-0.3em}} .

\bibitem[Lumley(1967)]{lumely1967}
{\sc \au{Lumley, J.~L.}} \yr{1967} The structure of inhomogeneous turbulent
  flows.  \bt{In {\em Atmospheric turbulence and radio wave propagation\/} (ed.
  \ed{A.~M. Yaglom \& V.~I. Tatarski})}.  \publ{Nauka}.

\bibitem[Maulik \& San(2017)]{MS2017}
{\sc \au{Maulik, R.} \& \au{San, O.}} \yr{2017}  \at{A neural network approach
  for the blind deconvolution of turbulent flows}.  \jt{J. Fluid Mech.}
  \bvol{831},  \pg{151--181}.

\bibitem[Milano \& Koumoutsakos(2002)]{Milano2002}
{\sc \au{Milano, M.} \& \au{Koumoutsakos, P.}} \yr{2002}  \at{Neural network
  modeling for near wall turbulent flow}.  \jt{J. Comput. Phys.}  \bvol{182},
  \pg{1--26}.

\bibitem[Murray {\em et~al.\/}(2009)Murray, Sallstrom \& Ukeiley]{Murray2009}
{\sc \au{Murray, N.}, \au{Sallstrom, E.} \& \au{Ukeiley, L.}} \yr{2009}
  \at{Properties of subsonic open cavity flow fields}.  \jt{Phys. Fluids}
  \bvol{21}~(9),  \pg{095103}.

\bibitem[Noack {\em et~al.\/}(2016)Noack, Stankiewicz, Morzynski, J. \&
  Schmid]{NSMS2016}
{\sc \au{Noack, B.~R.}, \au{Stankiewicz, W.}, \au{Morzynski, M.}, \au{J., P.}
  \& \au{Schmid, P.~J.}} \yr{2016}  \at{Recursive dynamic mode decomposition of
  transient and post-transient wake flows}.  \jt{J. Fluid Mech.}  \bvol{809},
  \pg{843--872}.

\bibitem[Omata \& Shirayama(2019)]{omata2019}
{\sc \au{Omata, N.} \& \au{Shirayama, S.}} \yr{2019}  \at{A novel method of
  low-dimensional representation for temporal behavior of flow fields using
  deep autoencoder}.  \jt{AIP Adv.}  \bvol{9}~(1),  \pg{015006}.

\bibitem[Prechelt(1998)]{prechelt1998}
{\sc \au{Prechelt, L.}} \yr{1998}  \at{Automatic early stopping using cross
  validation: quantifying the criteria}.  \jt{Neural Netw.}  \bvol{11}~(4),
  \pg{761--767}.

\bibitem[Roweis \& Lawrence(2000)]{Roweis2000}
{\sc \au{Roweis, S.} \& \au{Lawrence, S.}} \yr{2000}  \at{Nonlinear
  dimensionality reduction by locally linear embedding}.  \jt{Science}
  \bvol{290},  \pg{2323--2326}.

\bibitem[Rowley \& Dawson(2017)]{Rowley2017}
{\sc \au{Rowley, C.~W.} \& \au{Dawson, S. T.~M.}} \yr{2017}  \at{Model
  reduction for flow analysis and control}.  \jt{Annu. Rev. Fluid. Mech.}
  \bvol{49},  \pg{387--417}.

\bibitem[Salehipour \& Peltier(2019)]{SP2019}
{\sc \au{Salehipour, H.} \& \au{Peltier, W.~R}} \yr{2019}  \at{Deep learning of
  mixing by two `atoms' of stratified turbulence}.  \jt{J. Fluid Mech.}
  \bvol{861},  \pg{R4}.

\bibitem[Samimy {\em et~al.\/}(2007)Samimy, Debiasi, Caraballo, Serrani, Yuan,
  Little \& Myatt]{SAMIMY2007}
{\sc \au{Samimy, M.}, \au{Debiasi, M.}, \au{Caraballo, E.}, \au{Serrani, A.},
  \au{Yuan, X.}, \au{Little, J.} \& \au{Myatt, J.~H.}} \yr{2007}  \at{Feedback
  control of subsonic cavity flows using reduced-order models}.  \jt{J. Fluid
  Mech.}  \bvol{579},  \pg{315--346}.

\bibitem[San \& Maulik(2018)]{SM2018}
{\sc \au{San, O.} \& \au{Maulik, R.}} \yr{2018}  \at{Extreme learning machine
  for reduced order modeling of turbulent geophysical flows}.  \jt{Phys. Rev.
  E}  \bvol{97},  \pg{04322}.

\bibitem[Schmid(2010)]{Schmid2010}
{\sc \au{Schmid, P.~J.}} \yr{2010}  \at{Dynamic mode decomposition of numerical
  and experimental data}.  \jt{J. Fluid Mech.}  \bvol{656},  \pg{5--28}.

\bibitem[Srinivasan {\em et~al.\/}(2019)Srinivasan, Guastoni, Azizpour,
  Schlatter \& Vinuesa]{SGASV2019}
{\sc \au{Srinivasan, P.~A.}, \au{Guastoni, L.}, \au{Azizpour, H.},
  \au{Schlatter, P.} \& \au{Vinuesa, R}} \yr{2019}  \at{Predictions of
  turbulent shear flows using deep neural networks}.  \jt{Phys. Rev. Fluids}
  \bvol{4},  \pg{054603}.

\bibitem[Taira {\em et~al.\/}(2017)Taira, Brunton, Dawson, Rowley, Colonius,
  McKeon, Schmidt, Gordeyev, Theofilis \& Ukeiley]{taira2017}
{\sc \au{Taira, K.}, \au{Brunton, S.~L.}, \au{Dawson, S. T.~M.}, \au{Rowley,
  C.~W.}, \au{Colonius, T.}, \au{McKeon, B.~J.}, \au{Schmidt, O.~T.},
  \au{Gordeyev, S.}, \au{Theofilis, V.} \& \au{Ukeiley, L.~S.}} \yr{2017}
  \at{Modal analysis of fluid flows: An overview}.  \jt{AIAA J.}
  \bvol{55}~(12),  \pg{4013--4041}.

\bibitem[Taira {\em et~al.\/}(2019)Taira, Hemati, Brunton, Sun, Duraisamy,
  Bagheri, Dawson \& Yeh]{taira2019}
{\sc \au{Taira, K.}, \au{Hemati, M.~S.}, \au{Brunton, S.~L.}, \au{Sun, Y.},
  \au{Duraisamy, K.}, \au{Bagheri, S.}, \au{Dawson, S.} \& \au{Yeh, CA}}
  \yr{2019}  \at{Modal analysis of fluid flows: Applications and outlook}.
  \jt{{\rm arXiv:1903.05750}\hspace{-0.3em}} .

\end{thebibliography}

\end{document}